\documentclass[%
 reprint,
superscriptaddress,
 showpacs,preprintnumbers,
 amsmath,amssymb,
prb,
]{revtex4-1}

\usepackage{graphicx}% Include figure files
\usepackage{dcolumn}% Align table columns on decimal point
\usepackage{bm}% bold math
\begin{document}

%\preprint{APS/123-QED}

\title{Superconductivity in the noncentrosymmetric half-Heusler compound LuPtBi : \\
A possible topological superconductor}
% Force line breaks with \\

\author{F. F. Tafti} 
 \email{fazel.fallah.tafti@usherbrooke.ca}
\affiliation{D\'epartement de physique \& RQMP, Universit\'e de Sherbrooke, Sherbrooke, Qu\'ebec, Canada J1K 2R1}

\author{Takenori Fujii}
\affiliation{ Cryogenic Research Center, University of Tokyo, Bunkyo, Tokyo 113-0032, Japan }%

\author{A. Juneau-Fecteau}
\affiliation{D\'epartement de physique \& RQMP, Universit\'e de Sherbrooke, Sherbrooke, Qu\'ebec, Canada J1K 2R1}

\author{S. Ren\'{e} de Cotret}
\affiliation{D\'epartement de physique \& RQMP, Universit\'e de Sherbrooke, Sherbrooke, Qu\'ebec, Canada J1K 2R1}

\author{N. Doiron-Leyraud}
\affiliation{D\'epartement de physique \& RQMP, Universit\'e de Sherbrooke, Sherbrooke, Qu\'ebec, Canada J1K 2R1}

\author{Atsushi Asamitsu}
 \affiliation{ Cryogenic Research Center, University of Tokyo, Bunkyo, Tokyo 113-0032, Japan }%

\author{Louis Taillefer}
 \email{louis.taillefer@usherbrooke.ca}
\affiliation{D\'epartement de physique \& RQMP, Universit\'e de Sherbrooke, Sherbrooke, Qu\'ebec, Canada J1K 2R1}
\affiliation{Canadian Institute for Advanced Research, Toronto, Ontario, Canada M5G 1Z8}

%\affiliation{
%\affiliation{%


\date{\today}% It is always \today, today,

\begin{abstract}

We report superconductivity in the ternary half-Heusler compound LuPtBi, with $T_c=1.0$~K and $H_{c2}=1.6$~T.
The crystal structure of LuPtBi lacks inversion symmetry, hence the material is a noncentrosymmetric superconductor.
Magnetotransport data show semimetallic behavior in the normal state,
which is evidence for the importance of spin-orbit interaction.
Theoretical calculations indicate that the strong spin-orbit interaction in
LuPtBi should cause strong band inversion, making this material a promising candidate for 3D topological superconductivity.

\end{abstract}

\maketitle

%\tableofcontents

\section{\label{introduction} Introduction}

Half-Heusler ternary compounds attract increasingly more attention as new multifunctional materials with spintronic and thermo-electric applications.\cite{felser_spintronics:_2007, mastronardi_antimonides_1999,jung_thermoelectric_2001}
Their simple 111 stoichiometric composition, chemically formulated as XYZ, contains a lanthanide (X), a transition metal (Y), and either Sb or Bi (Z). 
Chemical substitution with different elements from the periodic table tunes the electronic structure of the final product to semiconducting,\cite{shan_fabrication_2012}
semimetallic,\cite{xia_electrical_2001}
heavy fermion,\cite{canfield_magnetism_1991,fisk_massive_1991}
or superconducting\cite{goll_thermodynamic_2008,butch_superconductivity_2011} behavior.  
Recent theoretical work presents these highly tunable compounds as new platforms for topological quantum phenomena due to the presence of strong spin-orbit interactions. \cite{chadov_tunable_2010,lin_half-heusler_2010, franz_topological_2010}

\begin{figure}[t]
\centering
\includegraphics[width=3.4 in]{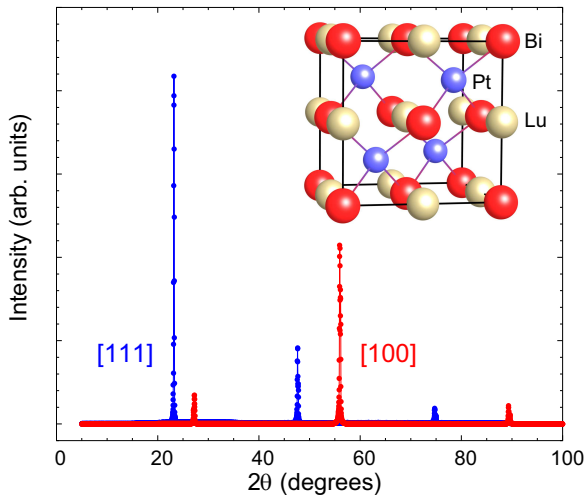}
\caption{\label{Fig1} X-ray diffraction data along the [100] (red) and [111] (blue) crystallographic directions. The inter-planar distance along the [111] direction is 3.801~\AA, consistent with the cubic lattice parameter $a=6.578$~\AA. Inset: the conventional unit cell of LuPtBi. This noncentrosymmetric structure, common to all the XYZ half-Heusler family, belongs to the $F\bar{4}3m$ space group. By taking out the X atom (Lu in this case), we obtain the zinc-blend structure of the HgTe. }
\end{figure}

The original theoretical prediction of the 2D topological insulators was based on the $\Gamma_6/\Gamma_8$ band inversion in the bulk of (Cd,Hg)Te quantum well due to strong spin-orbit coupling which gives rise to the conducting surface states.\cite{qi_topological_2011, bernevig_quantum_2006}
The same theoretical frame-work is readily extended to the half-Heusler compounds based on the similarity of their crystal structure to the zinc-blend structure of the (Cd,Hg)Te system.\cite{feng_half-heusler_2010}
Half-Heusler compounds crystallize in the space group $F\bar{4}3m$ composed of three FCC sub-lattices placed at X(0,0,0), Y(1/4,1/4,1/4), and Z(3/4,3/4,3/4) along the cubic diagonal (Fig \ref{Fig1}). 
The choice of XYZ elements determines the strength of spin-orbit  coupling which is proportional to the atomic number $z$.
%\begin{equation}
%\lambda_{SO} = \frac{e\textrm{g}\mu_B}{2mc^2} \frac{z}{r^3}
%\label{spinorbit}
%\end{equation}
Band structure calculations\cite{chadov_tunable_2010} have established a linear correlation between the total atomic number $z_{tot} = z_X+z_Y+z_Z$ in the XYZ composition of the half-Heuslers and the band inversion amplitude $|E_{\Gamma_6}-E_{\Gamma_8}|$.

%Topological order is a form of long range quantum entanglement which contrary to conventional Landau-type order, does not break any symmetries of the original Hamiltonian \cite{xiao_light_2004}. 
%Finding materials in which topological and conventional order coexist is extremely helpful in understanding the consequences of the topological order. 
%Such coexistence is suggested to realize in half-Heusler compounds due to their high electronic tunability  \cite{franz_topological_2010}. 

A choice of YZ = PtBi in particular favors strong spin orbit coupling due to the heavy mass of the atomic constituents. 
Amongst the XPtBi family, LaPtBi and YPtBi superconduct.\cite{goll_thermodynamic_2008,butch_superconductivity_2011, bay_superconductivity_2012}
Both compounds are in the band-inverted regime, hence their Cooper pairs are formed out of what is expected to be a topologically nontrivial band structure. 
$z_{tot}$ is maximal for X = Lu and so is the band inversion amplitude, hence LuPtBi is the most promising candidate for 3D  topological superconductivity amongst the half-Heusler compounds.\cite{chadov_tunable_2010, lin_half-heusler_2010}
We report the discovery of superconductivity at $T_c=1.0$ K in this promising material.

LuPtBi is interesting for a second reason.
%
%When the crystal structure respects inversion symmetry, Cooper pairing will be restricted to either singlet or triplet channels. 
Its noncentrosymmetric (NCS) crystal structure in which inversion symmetry is violated, 
implies that parity is no longer a conserved quantum number, and mixed singlet-triplet pairing is possible.\cite{saxena_superconductivity:_2004}
As a result, novel properties may follow, such as large Pauli-limiting fields and helical vortex states \cite{edelstein_magnetoelectric_1995, kaur_helical_2005}.
The helical vortex state is equivalent to the Fulde-Ferrell-Larkin-Ovchinnikov (FFLO) state in centrosymmetric superconductors, 
where Cooper pairs have finite center-of-mass momentum and the order parameter is spatially modulated along the field. 
The helical vortex state occurs within Ginzburg-Landau theory by including strong spin-orbit  coupling,\cite{samokhin_magnetic_2004}
a condition pertinent to LuPtBi.

%The article is organized as follows.
%In section II, we provide experimental details.
%In section III, we present our transport and magnetic susceptibility data. 
%We show that the normal state properties provide evidence for strong SO interactions in LuPtBi, in addition to our %expectations from the atomic physics arguments presented above (Eq. \ref{spinorbit}).
%We present evidence for bulk superconductivity by showing both resistivity and magnetic susceptibility data and extract the %upper critical field.
%In section IV, we discuss our results in the light of theoretical works on both the noncentrosymmetric and the topological %aspects of superconductivity in LuPtBi.

%------------------ FIGURE 1 --------------------------------------------------------------------------------------------------------

%------------------------------------------------------------------------------------------------------------------------------------

%%%%%%%%%%%%%%%%%%%%%%%%%%%%%%%%%%%%%%%%%%%%%%%%%%%%%%%%%%%%%%%%%%%%%%%%%%%%%%%%%%%%%%%%%%%%%%%%%%%%%%%%%%%%%%%%%%%%%%%%%%%%
%%%%%%%%%%%%%%%%%%%%%%%%%%%%%%%%%%%%%%%%%%%  NEW SECTION  %%%%%%%%%%%%%%%%%%%%%%%%%%%%%%%%%%%%%%%%%%%%%%%%%%%%%%%%%%%%%%%%%%
%%%%%%%%%%%%%%%%%%%%%%%%%%%%%%%%%%%%%%%%%%%%%%%%%%%%%%%%%%%%%%%%%%%%%%%%%%%%%%%%%%%%%%%%%%%%%%%%%%%%%%%%%%%%%%%%%%%%%%%%%%%%

\section{\label{methods} Methods}

Single crystals of LuPtBi were grown in Bi flux.
X-ray diffraction patterns along the [100] and [111] directions, presented in Fig. \ref{Fig1}, show no evidence of impurity phases. 
The lattice constant of the cubic structure $a=6.578$~\AA~is consistent with previous reports.\cite{canfield_magnetism_1991} 
Energy dispersive X-ray spectrometry gives atomic percentages $32.3:34.6:33.09$ for Lu:Pt:Bi, confirming the stoichiometric ratio of the chemical composition.  

Four-probe resistivity measurements were performed from 300 to 0.3 K in a Cambridge Magnetic Refrigerator (CMR).
The current was applied along the high-symmetry [100] direction, and the magnetic field in the [010] direction.
The Hall effect was measured by reversing the field and antisymmetrizing the data at $H=\pm10$~T.
AC susceptibility was measured with the mutual inductance method using a system of four coils and a lock-in detector. 
We used a drive field of amplitude 0.03 Oe and frequency 1 kHz.

%%%%%%%%%%%%%%%%%%%%%%%%%%%%%%%%%%%%%%%%%%%%%%%%%%%%%%%%%%%%%%%%%%%%%%%%%%%%%%%%%%%%%%%%%%%%%%%%%%%%%%%%%%%%%%%%%%%%%%%%%%%%
%%%%%%%%%%%%%%%%%%%%%%%%%%%%%%%%%%%%%%%%%%%  NEW SECTION  %%%%%%%%%%%%%%%%%%%%%%%%%%%%%%%%%%%%%%%%%%%%%%%%%%%%%%%%%%%%%%%%%%
%%%%%%%%%%%%%%%%%%%%%%%%%%%%%%%%%%%%%%%%%%%%%%%%%%%%%%%%%%%%%%%%%%%%%%%%%%%%%%%%%%%%%%%%%%%%%%%%%%%%%%%%%%%%%%%%%%%%%%%%%%%%

\section{\label{results} Results}

%------------------ FIGURE 2 --------------------------------------------------------------------------------------------------------

\begin{figure}[t]
\includegraphics[width=3.4 in]{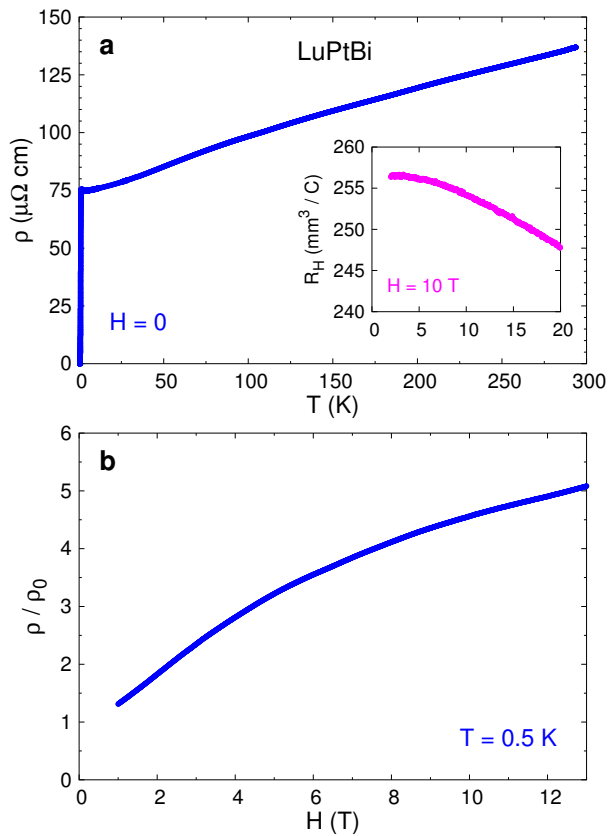}
\caption{\label{Fig2} (a) Electrical resistivity of LuPtBi as a function of temperature. 
Inset: Hall coefficient of LuPtBi as a function of temperature, below 20~K. 
(b) Normal-state resistivity of LuPtBi as a function of magnetic field, 
plotted as $\rho / \rho_0$ vs $H$ at $T=0.5$~K, with $\rho_0 = 74~\mu \Omega$~cm.}
\end{figure}

%------------------------------------------------------------------------------------------------------------------------------------

Fig. \ref{Fig2}(a) shows the temperature dependence of electrical resistivity in LuPtBi from 300 to 0.3 K.
The resistivity gradually decreases from $\rho=137$ $\mathrm{\mu\Omega\,cm}$ at room temperature to a residual value $\rho_0=74$ $\mathrm{\mu\Omega\,cm}$ at $T \to 0$.
The inset of Fig. \ref{Fig2}(a) shows the temperature dependence of the Hall coefficient $R_{\rm H}$ measured from $T=20$ to 2 K. 
At low temperatures, it saturates to a large positive value, $R_{\rm H}(0) = + 256$ mm$^{3}$/C,
corresponding to a hole concentration $n_{\rm H}=1/eR_{\rm H}=2.44\times10^{19}$ cm$^{-3}$, in a simple one-band model. 
Our resistivity and Hall data are comparable to a previous study of LuPtBi which report $\rho_0=105$ $\mathrm{\mu\Omega\,cm}$ and  $R_{\rm H}(0) = + 143$ mm$^{3}$/C hence $n_{\rm H}=1/eR_{\rm H}=4.3\times10^{19}$ cm$^{-3}$. \cite{mun2010} 
This study stops at $T=2$ K hence superconductivity has not been revealed.
Using the result of the one-band Drude model $\sigma_0=ne^2\tau/m$ and assuming a spherical Fermi surface where $k_{\rm F}=(3 \pi^2 n)^{1/3}$, we obtain a large mean free path $l=1.3$ $\mathrm{\mu m}$, confirming high sample quality.

Fig. \ref{Fig2}(b) shows the field dependence of the resistivity at $T=0.5$ K from $H=0$ to 13 T.
A strong positive magnetoresistance (MR) is observed, whereby $\rho$ increases by a factor 5 in 13 T.
% via $\text{MR}=\frac{\rho(13\text{T})-\rho(1\text{T})}{\rho(1\text{T})}=2.9$.
Our observation of a large mean free path is consistent with a large orbital MR.
Such high MR values and low carrier concentrations
%and weak temperature dependence of resistivity 
are typical characteristics of semimetals.\cite{pippard_magnetoresistance_2009}
 
Our use of a one-band Drude model to determine $n_H$ and $l$ is clearly naive, considering that semimetals are typically multi-band systems, nevertheless, we use these simple calculations for a first estimate of the physical parameters of the material.
We use the same model to calculate similar physical properties of the other two superconduting members of the half-Heusler series, namely YPtBi and LaPtBi, and compare them to LuPtBi in Table \ref{materials}.
The three compounds have comparable $T_c$ values and small concentration of carriers, comparable to degenerate semiconductors. \cite{schooley_dependence_1965}

%, comparable to degenerate semiconductors, which typically have much smaller $T_c$ values. \cite{schooley_dependence_1965}
%For example, $n$-doped SrTiO$_3$ at the same doping levels has $T_c=0.1$~K {\bf [REF?]}.

\begin{figure}[t]
\includegraphics[width=3.4 in]{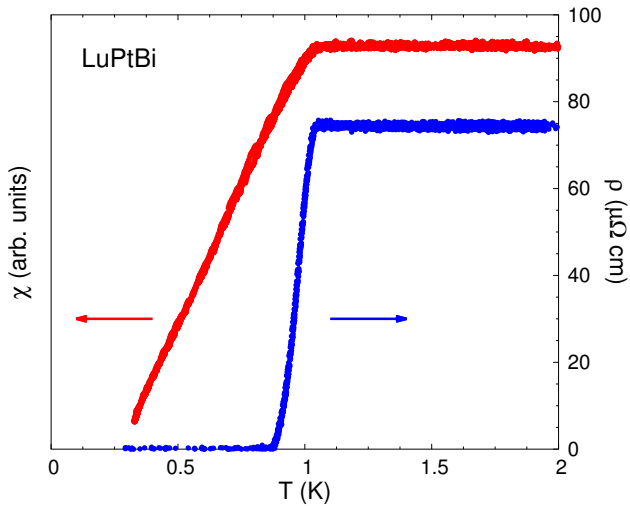}
\caption{\label{Fig3} Superconducting phase transition in LuPtBi, occurring at $T_c=1.0$~K. 
The transition is observed in both the resistivity (blue curve, right axis) and the magnetic susceptibility (red curve, left axis).}
\end{figure}

Fig. \ref{Fig3} shows the superconducting transition at $T_c=1.0~\pm~0.1$~K.
The transition is observed as a drop in both electrical resistivity and magnetic susceptibility, confirming bulk superconductivity in LuPtBi. 
%At low temperature, the resistivity is almost temperature independent, allowing for a straightforward extrapolation to $T=0$,
%giving a residual resistivity $\rho_0=74$~$\mathrm{\mu\Omega\,cm}$ (Fig. \ref{Fig3}).
The shape of the transition in AC susceptibility is similar to what has been observed in YPtBi.\cite{butch_superconductivity_2011}
%We observe a sharper resistivity drop in our LuPtBi single crystals compared to YPtBi which could be simply due to higher sample quality. 

%------------------ FIGURE 3 --------------------------------------------------------------------------------------------------------

%------------------------------------------------------------------------------------------------------------------------------------

 %------------------ TABLE  I --------------------------------------------------------------------------------------------------------
 
\begin{table}[b]%The best place to locate the table environment is directly after its first reference in text
\caption{\label{materials} Physical properties of the three superconducting members of the XPtBi series, with X = Y,\cite{bay_superconductivity_2012, butch_superconductivity_2011}
La,\cite{goll_thermodynamic_2008}
and Lu \cite{mun2010}. $T_c$ is defined as the onset of the superconducting drop in the resistivity vs $T$ curve; $H_{c2}$ is extracted from the temperature dependence of the onset of the superconducting drop in the resistivity vs $H$ curves; the Hall concentration $n_{\rm H}=1/eR_{\rm H}(0)$. 
Note the different values for the low-temperature $n_{\rm H}$ of YPtBi obtained from two different studies. }

\begin{ruledtabular}
\begin{tabular}{ccccc}
X &
$T_c ~ \mathrm{(K)}$ &
$H_{c2} ~ \mathrm{(T)}$ &
$n_{\rm H} ~ \mathrm{(cm^{-3})}$ &
Reference\\
\colrule
\\
La & 0.9 & 1.5 & $6\times 10^{18}$ & Ref.~\onlinecite{goll_thermodynamic_2008} \\
Y & 0.8 & 1.5 & $2\times 10^{18}$ & Ref.~\onlinecite{butch_superconductivity_2011} \\
Y & 0.8 & 1.2 & $2\times 10^{19}$ & Ref.~\onlinecite{bay_superconductivity_2012}  \\ 
Lu & - & - & $4\times 10^{19}$ & Ref.~\onlinecite{mun2010}  \\ 
Lu & 1.0 & 1.6 & $2\times 10^{19}$ & this work \\
\end{tabular}
\end{ruledtabular}
\end{table}  
		
%------------------------------------------------------------------------------------------------------------------------------------

\begin{figure}
\includegraphics[width=3.4 in]{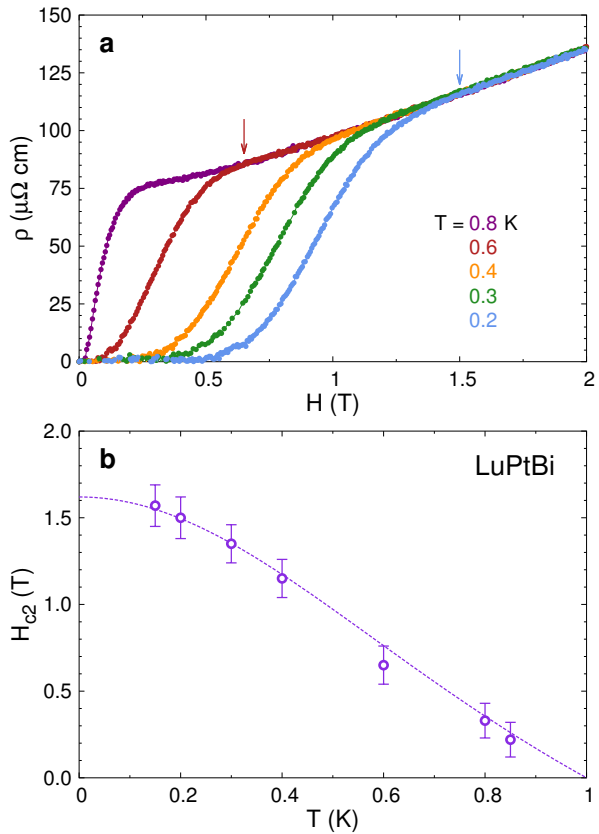}
\caption{\label{Fig4} (a) Field dependence of the resistivity in LuPtBi, at different temperatures. 
A selected number of isotherms are shown as indicated. 
$H_{c2}(T)$ is taken as the full recovery of the normal-state resistivity,
as marked by arrows for $T = 0.2$~K and 0.6~K.
(b) $H_{c2}$ as a function of temperature. 
The dotted line is a fit to the Ginzburg-Landau expression (Eq.\ref{GL}), which yields $H_{c2}(0)=1.6$~T.}
\end{figure}

We have determined the upper critical field $H_{c2}$ of LuPtBi by studying the field dependence of the resistivity at different temperatures.
Fig. \ref{Fig4}(a) shows resistivity curves as a function of field from $T=0.2$~K to 0.8~K. 
$H_{c2}(T)$ at each temperature is defined as the full recovery of the normal-state resistivity.
Fig. \ref{Fig4}(b) shows the temperature dependence of $H_{c2}(T)$ extracted from the data in Fig. \ref{Fig4}(a).
We evaluate the zero temperature limit of the upper critical field to be $H_{c2}(0)=1.6 \pm 0.1$~T by fitting our data to the generalized Ginzburg-Landau model:

\begin{equation}
H_{c2}(T)=H_{c2}(0)\frac{1-t^2}{1+t^2}~~~~~,
\label{GL}
\end{equation}

where $t=T/T_c$ (Fig. \ref{Fig4}b).
Using the zero-temperature relation $H_{c2} = \phi_0 / 2\pi \xi_0^2$, we extract the coherence length $\xi_0=14$ nm.
Comparing the coherence length with the mean free path we find our sample satisfying the clean limit condition $l\gg \xi_0$.

%------------------ FIGURE 4 --------------------------------------------------------------------------------------------------------

%--------------------------------------------------------------------------------------------------------

%%%%%%%%%%%%%%%%%%%%%%%%%%%%%%%%%%%%%%%%%%%%%%%%%%%%%%%%%%%%%%%%%%%%%%%%%%%%%%%%%%%%%%%%%%%%%%%%%%%%%%%%%%%%%%%%%%%%%%%%%%%%
%%%%%%%%%%%%%%%%%%%%%%%%%%%%%%%%%%%%%%%%%%%  NEW SECTION  %%%%%%%%%%%%%%%%%%%%%%%%%%%%%%%%%%%%%%%%%%%%%%%%%%%%%%%%%%%%%%%%%%
%%%%%%%%%%%%%%%%%%%%%%%%%%%%%%%%%%%%%%%%%%%%%%%%%%%%%%%%%%%%%%%%%%%%%%%%%%%%%%%%%%%%%%%%%%%%%%%%%%%%%%%%%%%%%%%%%%%%%%%%%%%%

\section{\label{discussion} Discussion}

According to transport experiments, depending on the choice of the rare earth element X, the XPtBi compounds may be either semiconducting or semimetallic.\cite{canfield_magnetism_1991}
The former is favoured by the lighter rare earth atoms while the latter is favoured by the heavier ones.
Band structure calculations show that in the absence of spin-orbit  interaction, XPtBi is a semiconductor.
Semimetallic properties appear only when the spin-orbit  interaction is included.\cite{oguchi_electronic_2001}
Our observations of the weak temperature dependence of resistivity, the small concentration of carriers, 
and the large magnetoresistance in LuPtBi indicate that the normal state is a semimetal, hence spin-orbit  interaction must play a significant role.
A large spin-orbit  coupling is also expected from atomic physics considerations since Lu has the largest atomic number amongst the lanthanides.

LuPtBi is an unconventional superconductor in two respects. 
First, it is a noncentrosymmetric superconductor, because its crystal structure lacks inversion symmetry.
Secondly, superconductivity in the bulk of the material emerges from a band structure which is likely to be topologically nontrivial. 
Below, we  discuss both aspects in turn.

%------------------ TABLE  II   --------------------------------------------------------------------------------------------------------

\begin{table}[t]
%The best place to locate the table environment is directly after its first reference in text
\caption{\label{NCS} Transition temperature and characteristic field scales of some NSC superconductors,
 including three heavy-fermion systems \cite{bauer_heavy_2004, sugitani_pressure-induced_2006, kimura_pressure-induced_2005} (top three lines) 
 and systems without $f$  electrons \cite{togano_superconductivity_2004, badica_superconductivity_2005, bay_superconductivity_2012} (bottom three lines). 
 The former group satisfies $H_{c2}>H_{\rm P}$, hence superconductivity is not Pauli limited,
 whereas the latter group satisfies $H_{\rm orb}<H_{c2}<H_{\rm P}$, hence superconductivity is Pauli limited. 
 LuPtBi belongs  to the latter group.}

\begin{ruledtabular}
\begin{tabular}{ccccc}
Material&
$T_c$~(K)&
$H_{\rm orb}$~(T)&
$H_{\rm P}$~(T)&
$H_{c2}$~(T) \\
\colrule
\\
CePt$_3$Si & 0.75 & 4.6 & 1.4 & 5.0 \\
CeIrSi$_3$ & 1.6 & 13.1 & 1.9 & 11.1 \\
CeRhSi$_3$ & 1.0 & 8.7 & 1.8 & 7.0 \\
\\
Li$_2$Pd$_3$B & 7.0  & 5.0  & 13  & 5.5  \\
Li$_2$Pt$_3$B & 2.7  & 1.0  & 5.0  & 2.0  \\
YPtBi & 0.8  & 1.0  & 1.4  & 1.2  \\
\end{tabular}
\end{ruledtabular}
\end{table} 

%------------------------------------------------------------------------------------------------------------------------------------

\emph{Noncentrosymmetic superconductivity}.
Superconductivity in a NCS system was first observed in the heavy-fermion metal CePt$_3$Si.\cite{bauer_heavy_2004}
Soon after, similar $f$-electron systems were discovered such as CeIrSi$_3$ and CeRhSi$_3$, both superconducting under pressure.\cite{sugitani_pressure-induced_2006, kimura_pressure-induced_2005}
NCS superconductivity has also been discovered in non-$f$ systems such as Li$_2$Pd$_3$B and Li$_2$Pt$_3$B.\cite{togano_superconductivity_2004, badica_superconductivity_2005}

In the absence of a center of inversion, an asymmetric crystal field potential creates an electric field $\vec{E}=-\nabla\Phi$ which can generate a Rashba spin-orbit interaction $(\vec{E}\times\vec{p})\cdot\vec{S}$. 
This interaction splits the Fermi surface and introduces a certain helicity to the electrons on each surface hence pure spin-singlet or spin-triplet pairings can no longer be valid descriptions of the pairing state.
Mixed singlet-triplet pairing is one intriguing possibility. 

%One exciting finding in the discovery of the NCS heavy fermion superconductors, was that their $H_{c2}$ would exceed the Pauli limiting field $H_P$ which was taken as evidence for spin-triplet pairing.
%As discussed above, pure spin-triplet pairing is not necessarily a correct description for these systems.
%In fact, the $H_{c2}$ of the NCS superconductors which do not have $f$-electrons such as Li$_2$(Pd,Pt)$_3$B was found to be less than $H_P$.

We evaluate $H_{c2}=1.6$~T in LuPtBi from a generalized Ginzburg-Landau analysis (Fig. \ref{Fig4}(a)).
Using the Werthamer-Helfand-Hohenberg formula in the clean limit $H_{\rm orb}=0.72T_c\left[-dH_{c2}/dT\right]_{T_c}$, we evaluate the orbital limiting field $H_{\rm orb}=1.24$ T.
Using $H_{\rm P}=\Delta/\sqrt{2}\mu_{\rm B}$ and $\Delta=1.76~k_{\rm B} T_c$, we evaluate the Pauli limiting field $H_P=1.85$ T.
Since $H_{\rm orb}<H_{c2}<H_{\rm P}$, superconductivity in LuPtBi is Pauli limited.
The Maki parameter $\alpha =  \sqrt{2}H_{\rm orb}/H_{\rm P}$ is less than one hence the FFLO state is not favorable.\cite{gruenberg_fulde-ferrell_1966}

Table \ref{NCS} summarizes $T_c$, $H_{c2}$, $H_{\rm orb}$, and $H_{\rm P}$ for a number of NCS superconductors, including the heavy fermion systems which contain free $f$-electrons and the ones with no $f$-electrons. 
$H_{c2}$ of the heavy-fermion NCS superconductors clearly exceeds the Pauli limiting field $H_{\rm P}$, 
suggestive of triplet pairing.
In the non-$f$ NCS superconductors $H_{c2}<H_{\rm P}$,
hence spin triplet pairing is not an obvious possibility.
A recent careful study of YPtBi shows that the temperature dependence of $H_{c2}$ at different pressures collapses onto a single universal curve different from the standard curve expected from spin-singlet superconductors, hence triplet pairing is not entirely ruled out.\cite{bay_superconductivity_2012}
Our data shows that similar to YPtBi and the other non-$f$ systems, $H_{c2}<H_{\rm P}$ in LuPtBi.
This similarity between LuPtBi and the non-$f$ systems is not surprising, since the $f^{14}$ shell of Lu$^{3+}$ is full and does not contribute to the electric conduction.   
The fact that $H_{c2}>H_{\rm P}$ only for heavy-fermion NCS superconductors, raises the question: 
what is the effect of strong correlations in determining the magnitude of $H_{c2}$ and the pairing symmetry in the NCS superconductors?
Further theoretical work is needed to answer that question.

\emph{Topological superconductivity}.
LuPtBi is the most promising candidate for 3D topological superconductivity amongst the half-Heusler series because of its maximal band inversion strength.\cite{feng_half-heusler_2010,chadov_tunable_2010,al-sawai_topological_2010}
The other two superconducting compounds in this series, YPtBi and LaPtBi, include rare-earth ions with much smaller atomic numbers and no $f$-electrons.
Lu$^{3+}$, contrary to the Y$^{3+}$ and La$^{3+}$, has a full $f$-shell and a much larger $z$ number.
Our finding in LuPtBi shows that superconductivity is a common trend in the half-Heusler systems where $f$-electrons do not contribute to the conduction band.
%Band structure calculations have established a linear correlation between the total atomic number $z_{tot} = z_X+z_Y+z_Z$ in the XYZ composition of the half-Heuslers and the and the band inversion amplitude $|E_{\Gamma_6}-E_{\Gamma_8}|$ which puts LuPtBi on top of the list of the candidates for 3D topological superconductivity in this series \cite{feng_half-heusler_2010,chadov_tunable_2010,al-sawai_topological_2010}.

In topological insulators an inverted band gap separates the $\Gamma_6$ ($s$ orbital) and the $\Gamma_8$ ($p$ orbital) bands, with the latter being above the former, in reverse order to the trivial insulators. 
This band inversion inevitably causes a band crossing at the surface of the material, giving rise to conducting surface states which are helical due to spin-orbit interaction. 
In the bulk of a topological superconductor, the band gap of the topological insulator is replaced by a superconducting gap. 
At the surface of a topological superconductor, the conventional electrons which form the helical edge states in the topological insulator are replaced by Majorana fermions. \cite{hasan_colloquium:_2010,moore_birth_2010, qi_time-reversal-invariant_2009} 
A major incentive in the search for Majorana fermions is their potential application to topological quantum computing.\cite{kitaev_fault-tolerant_2003,nayak_non-abelian_2008}  
%Due to their Non-Alebian statistics, Majorana fermion are exceptionally well protected from local sources of decoherence which makes them ideal for fault free quantum computations \cite{kitaev_fault-tolerant_2003,nayak_non-abelian_2008}.  
%Majorana fermions can be achieved in two ways: 
%(a) By a proximity effect at the interface between a topological insulator (TI) and an ordinary superconductor (OSC), the metallic surface states of the TI become superconducting and where the vortices of the OSC cross the interface, a zero-energy Majorana fermion will be trapped in the core of the vortex \cite{fu_superconducting_2008}.  
%(b) If superconductivity coexists with TI in the bulk of a material, for example in the case of Cu$_x$Bi$_2$Se$_3$, Mjoarana bound states can exist on the surface without dissipating into the bulk of the material as long as the insulating gap of the TI is large enough \cite{hosur_majorana_2011}.   

In conclusion, LDA calculations show that LuPtBi is electronically tuned to a
%an all bands touching 
semimetallic state from a parent topological-insulator state\cite{lin_half-heusler_2010}
and our transport data reveal semimetallic behavior in LuPtBi.
%In the language of Fu-Kane classification, the band parity product $Z_2=-1$ hence a strong TI\cite{fu_topological_2007}.
%Verifying the topological character of a superconductor is not a trivial task because topological order is not a broken symmetry state with an experimentally detectable order parameter. 
%One way to get around this problem is to design materials which can accommodate both kinds of order, namely the symmetry-breaking order and the topological order. 
%
Our finding of superconductivity
%, a state with broken gauge symmetry, 
in a material that satisfies the requirements of a topologically nontrivial band structure
%is of great importance.
offers the exciting possibility of a 3D topological superconductor.

We thank K. Behnia, J. Paglione, K. Samokhin and S.-C. Zhang for helpful discussions.
The work at Sherbrooke was supported by a Canada Research Chair, CIFAR, NSERC, CFI, and FQRNT.
%

%\nocite{*}

\bibliography{apssamp}% Produces the bibliography via BibTeX.

\end{document}